\documentstyle[11pt,newpasp]{article}

\begin{document}

\title{Spiral Galaxies with HST - Near-Infrared Properties of Bulges}
\author{M. Seigar}
\affil{Sterrenkundig Observatorium, Universiteit Gent, Krijgslaan 281-S9, B-9000 Gent, Belgium}

\begin{abstract}
We have obtained {\tt HST-NICMOS} images for 72 nearby spiral galaxy bulges in the H-band. These data show that galaxies with regular bulges have steeper nuclear cusps than galaxies with irregular bulges. We also show than galaxies with regular bulges fall on the same part of the ($\langle \gamma \rangle$, magnitude)-diagram as elliptical galaxies. This implies that early-type spiral galaxy bulges have a formation process in common with ellipticals.
\end{abstract}

\section{Introduction}

The nuclei of galaxies harbor clues to the processes involved in the formation of the host galaxies. Although, extensive work has been performed using HST imaging of nearby elliptical galaxies (e.g., Lauer et al. 1995; Carollo et al. 1997a, b; Faber et al. 1997), there is a lack of work concerning the formation of disk galaxies. There are a few recent studies that have started to address this. Peletier \& Balcells (1999) published results obtained using multi-color observations with {\tt NICMOS} and {\tt WFPC2} data of 20 nearby early-type galactic bulges. They found that the nuclei were typically dusty with a small age spread among bulges of early-type spiral galaxies (about 2-3 Gyr). There has also been an optical study by Carollo et al. (1997c, 1998) and Carollo \& Stiavelli (1998). Here we present NICMOS data of the galaxies that they observed.

\section{Analytical fits to the nuclear surface brightness profiles}

The nuclear surface brightness profiles of early-type galaxies is described by Lauer et al. (1995). Their description was used by Carollo et al. (1997b, 1998) to describe nuclear profiles of spiral galaxies imaged with {\tt WFPC2} and we now use it here to describe the profiles of 72 galaxies imaged with {\tt NICMOS} on board {\tt HST} in the H-band. There is an uncertainty that exists in removing the nuclear compact source contribution from the galaxy light profile, as the form of the light profile of the compact source is not well defined. This uncertainty was quantified by performing several different light profile fits to the same galaxy with a variable inner radial cutoff. Of the 72 galaxies, 55 were fitted. From these fits the average logarithmic nuclear slope $\langle \gamma \rangle$ has been derived.

\begin{figure}
\includegraphics{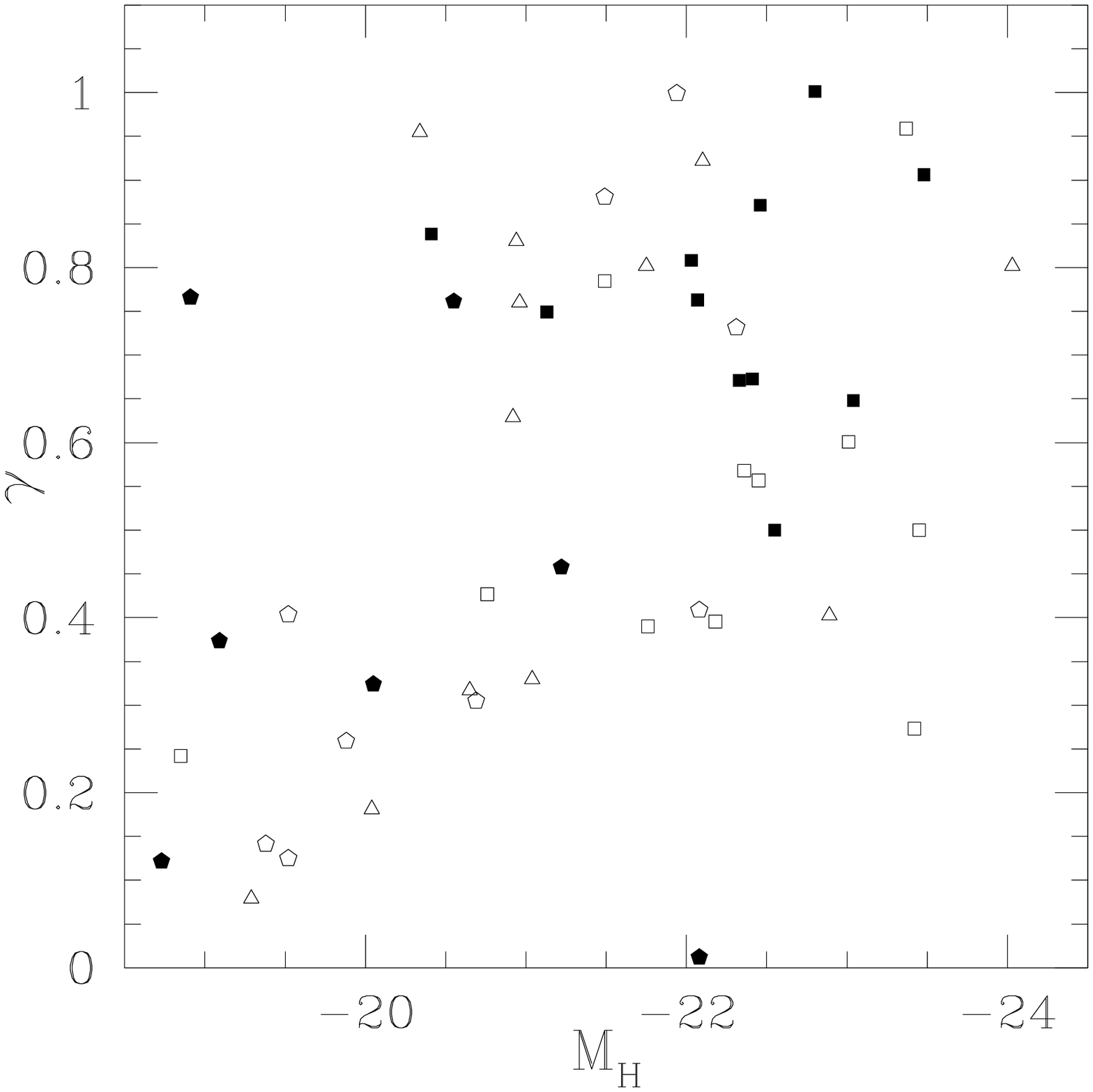}
\includegraphics{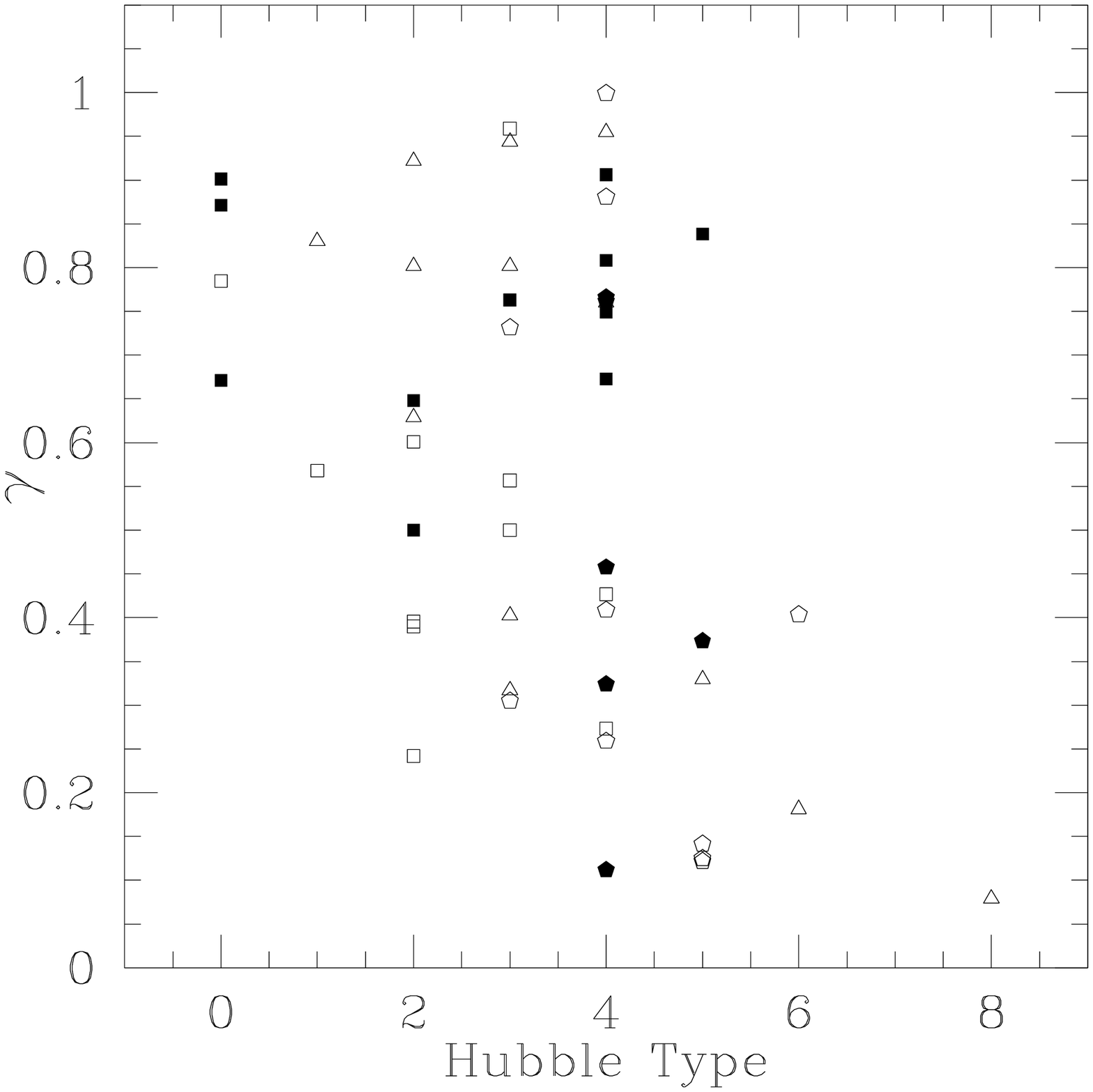}
\vspace{4.5cm}
\caption{The left hand plot shows average nuclear slope $\langle \gamma \rangle$ inside $0.1^{\prime\prime}-0.5^{\prime\prime}$ versus absolute {\tt F160W} magnitude of the spheroidal component. The right hand plot shows average nuclear slope inside $0.1^{\prime\prime}-0.5^{\prime\prime}$ ($\langle \gamma \rangle$) versus Hubble type. Hollow pentagons are single exponential galaxies, solid pentagons double exponential galaxies, hollow squares $R^{1/4}$ law bulges, solid squares $R^{1/4}$ plus exponential bulges, and triangles those galaxies with no acceptable fit.}
\end{figure}

Figure 1 shows that galaxies fitted with $R^{1/4}$ law profiles are generally found in the top right hand part of the ($\langle \gamma \rangle$, $M_H$) diagram, with exponential law fits being in the bottom left hand area, coincident with the area where elliptical galaxies are usually found (Lauer et al. 1995; Carollo et al. 1997a; Faber et al. 1997; Carollo \& Stiavelli 1998). Also a clear dichotomy is shown with early-type spirals generally have steeper cusps than late-type spirals, as well as a trend for galaxies with $R^{1/4}$ law fits to have steeper cusps than those with exponential law light-profile fits. Both of these results are in agreement with the main results of the optical study performed by Carollo \& Stiavelli (1998).

\section{Conclusions}

Our main result is that $R^{1/4}$-law bulges and exponential bulges have significantly different nuclear stellar cusps. Specifically, $R^{1/4}$ law bulges have steep stellar cusps which steepen with decreasing luminosity. Their stellar cusp slopes are also comparable to those of elliptical galaxies of similar luminosities. By contrast, in exponential bulges the inward extrapolation performed imply shallow cusp slopes. This is similar to the results of optical studies (Carollo \& Stiavelli 1998).

\end{document}